\documentclass[twocolumn,runningheads]{svjour3}

\usepackage[latin1]{inputenc}
\usepackage{times}

\usepackage{psfrag,amsmath, amssymb, amscd, amsfonts, latexsym, epsf, graphicx,amscd}
\usepackage{natbib}
\usepackage{breqn}
\usepackage{subcaption}
\usepackage[switch]{lineno}

\def\bbbr{{\mathbb R}}

\newcommand{\orth}{\bot}

\def\norm{\scriptsize\mbox{norm}}

\journalname{arXiv preprint}

\begin{document}


\titlerunning{The generalized Gaussian derivative model for visual receptive fields}
\title{\bf An axiomatic theory of spatial and spatio-temporal
  receptive fields with associated predicted variabilities under geometric image transformations%
\thanks{The support from the Swedish Research Council 
              (contract 2022-02969) is gratefully acknowledged. }}
\author{Tony Lindeberg}

\institute{Computational Brain Science Lab,
        Department of Computational Science and Technology,
        KTH Royal Institute of Technology,
        SE-100 44 Stockholm, Sweden.
        \email{tony@kth.se}.
      ORCID: 0000-0002-9081-2170.}

\date{}

\maketitle

\begin{abstract}
  \noindent
  Receptive fields in the primary visual cortex vary in both their
  shape and their selectivity properties, but the principles underlying
  these variabilities remain an important question regarding models of
  vision. Here, we review main components in a normative
  theory for visual receptive fields, that relates such variabilities
  to the influence of the geometric image transformations induced
  by varying the viewing conditions between different views
  of the same object or spatio-temporal event. The theory predicts
  that the responses from simple cells from different views can be
  matched, by transforming the shape parameters of the receptive
  fields, provided that the receptive fields are covariant under
  locally linearized geometric image transformations.
  This motivates variabilities in the receptive field
  shapes along the degrees of freedom of the geometric image
  transformations.
  These results are consistent with
  observed variabilities in orientation, direction and speed
  selectivity properties for higher mammals, as well as with
  observed variabilities in the degree of elongation for mice.
Together these results
  provide a mathematically based framework for how diversity in receptive
  field shapes could support matching of visual information under
  variations in viewing conditions and relative motion
  between objects/events in the environment and the observer.
  
  \keywords{Visual receptive field \and Variability \and Primary visual cortex \and
    Gaussian derivative model \and Geometric transformations 
    \and Theoretical neuroscience}
\end{abstract}

\section{Introduction}
\label{sec-intro}

The receptive fields in the primary visual cortex can aptly be
regarded as a visual front-end on which the mammalian vision system is
based. Understanding the computational functions of the receptive
fields of simple cells in the primary visual cortex
is, hence, fundamental for developing an understanding of the vision
system as well as for building integrated models of its neural functions.

A particularly important component in this context concerns the
variabilities in the shapes of the receptive fields that can be
observed in neurophysiological measurements. Are these variabilities
just side effect of the combined evolution and learning mechanisms
that determine the formation of the receptive fields, or could there
be underlying principles that determine the formation of the receptive
fields?

The subject of this paper is to give an overview of main results
from a normative theory of visual receptive fields
(Lindeberg \citeyear{Lin21-Heliyon}). This theory determines
the shapes of the receptive fields in an axiomatic manner, by starting
from a combination of symmetry properties,
using an approach similar to the way that theories in physics can be
derived.

According to this theory, purely spatial receptive fields are defined
as spatial derivatives of affine Gaussian kernels, whereas joint
spatio-temporal receptive fields are defined as spatio-temporal
derivatives of velocity-adapted affine Gaussian kernels.
The shapes of the spatial receptive fields are parameterized in
terms of a rich set of variabilities over
(i)~spatial scale, (ii)~preferred orientation in image space,
(iii)~degree of elongation, and
(iv)~a fourth spatial degree of freedom.
Joint spatio-temporal receptive fields are additionally also
parameterized over
(v)~preferred direction in joint space-time,
(vi)~preferred speed, and
(vii)~temporal scale.

These variabilities are theoretically derived from the assumption that
it should be possible to match the receptive field responses from
different views of the same object or spatio-temporal event.
The views may differ in 
(i)~the distance between the object/event and the observer,
(ii)~the viewing direction in relation to the local surface normals,
(iii)~the relative motion between the object/event and the observer,
and
(iv)~the speed by which spatio-temporal event occurs, either faster or
slower
(Lindeberg \citeyear{Lin26-JCompNeuroSci}).
Specifically, transformation properties have been derived concerning
the internal parameters of the either purely spatial or joint
spatio-temporal receptive fields.
These transformation properties make it possible to match the 
receptive field responses between multiple views of the
same object/event, using locally linearized,
first-order approximations%
\footnote{For the validity of the local linear approximations, we
  assume that the geometry varies slowly in relation to the
  spatio-temporal scales defining the support region of the
  receptive field. If not, the analysis should be performed
  at finer scales. Additionally, we exclude discontinuities and
  occlusions. See Lindeberg~(\citeyear{Lin26-JCompNeuroSci})
  Section~7 for a more extensive treatment regarding the validity
  of the model.}
of the otherwise general non-linear
projections (Lindeberg \citeyear{Lin25-JMIV}).

A direct prediction from this theory is that there ought to be
variabilities in the shapes of the receptive fields corresponding to
the degrees of freedom over geometric image transformations
(Lindeberg \citeyear{Lin25-BICY,Lin26-JCompNeuroSci}).
These predictions are specifically consistent with variabilities in
orientation, direction and speed selectivity properties of visual neurons
(Lindeberg \citeyear{Lin25-JCompNeurSci-spanelong,Lin25-JCompNeurSci-orisel,Lin25-PONE,Lin25-arXiv-dirsel}),
and are also consistent with variabilities in the degree of elongation
as have been observed in reconstructed receptive fields from
calcium imaging of neural responses in mice
(Meshcherina {\em et al.\/} \citeyear{MesAufLin20-bioRxiv}).
In this way, the generalized Gaussian derivative theory provides
a theoretical framework for relating mathematical vision
models to empirical cortical dynamics

The idealized models for the either spatial or joint spatio-temporal
receptive fields can furthermore be described by evolution properties
over spatial and temporal scales in terms of diffusion equations and
recurrence equations (Lindeberg \citeyear{Lin21-Heliyon,Lin26-JMIV}).
Thereby, the computations needed to implement
the idealized models of the visual receptive fields can be directly
mapped to diffusion and recurrence relations over neural
architectures. In this way, the generalized Gaussian
derivative theory provides a link between
compact functional theoretical models of visual receptive fields
and more fine-grained neural models thereof.

\section{Theoretical foundations}

When to formulate an axiomatic theory for visual receptive fields,
one may ask what assumptions one should start from.
In the area of scale-space theory, as developed by
Iijima (\citeyear{Iij62}), Koenderink (\citeyear{Koe84}),
Koenderink and van Doorn (\citeyear{KoeDoo92-PAMI}),
Lindeberg (\citeyear{Lin93-Dis,Lin96-ScSp}, \citeyear{Lin10-JMIV}) and
Weickert {\em et al.\/}\ (\citeyear{WeiIshImi99-JMIV}),
this problem has been addressed in a principled manner.
Following the axiomatic scale-space formulation in
Lindeberg (\citeyear{Lin10-JMIV}), which comprises both purely
spatial and joint spatio-temporal image domains, the basic
assumptions are, beyond certain formal regularity assumptions
in terms of integrability and strong continuity across scales:
\begin{itemize}
\item
  linearity, implying that as few irreversible decisions as
  possible should be made in the earliest processing stages,
\item
  translation covariance, meaning that all spatial and temporal
  points should be treated in the same manner,
\item
  semi-group property over scale, implying that the transformation
  from finer to coarser scales should always be of the same form, and
\item
  non-creation of structure from finer to coarser levels of scale $s$, in
  the sense that the value at any local maximum must not increase
  and that the value at any local minimum must not decrease.
\end{itemize}
Based on these assumptions, it is shown in Lindeberg
(\citeyear{Lin10-JMIV} Theorems~5--6), that any non-causal either purely spatial or
joint spatio-temporal smoothing kernel must satisfy
an evolution equation over the scale parameter $s$
in terms of a diffusion equation of the form
\begin{equation}
  \label{eq-vel-adapt-scsp-diff-eq}
  \partial_s L 
  = \frac{1}{2} \, \nabla^T \left( \Sigma \, \nabla L \right) - \delta^T \nabla L
\end{equation}
where $\nabla = (\partial_{x_1}, \partial_{x_2})^T$ or
$\nabla = (\partial_{x_1}, \partial_{x_2}, \partial_t)^T$
denotes the differentiation operator over either space
$x = (x_1, x_2)^T$ or joint space-time $p = (x_1, x_2, t)^T$
and $\Sigma$ denotes a positive definite either purely spatial or joint
spatio-temporal covariance matrix that determines the
amount of smoothing in the different direction of the
either purely spatial or joint spatio-temporal domain.
Over a joint spatio-temporal domain, the parameter
$\delta$ models Galilean drift over time. Over a pure
spatial domain, we do, however, disregard this parameter.

\section{Resulting idealized models for simple cells}

Complementing the above theory with the notion of explicit either
purely spatial or joint spatio-temporal derivatives, according to
the thereupon formulated normative theory for visual receptive fields
in Lindeberg (\citeyear{Lin21-Heliyon}),  does, via the fact that
convolution with the affine Gaussian kernel describes the solution
to the purely spatial affine diffusion equation
(\ref{eq-vel-adapt-scsp-diff-eq}) for $\delta = 0$,
lead to idealized models for simple cells formulated
in terms of affine Gaussian derivatives of the form
\begin{multline}
  \label{eq-spat-RF-model}
  T_{\text{simple}}(x_1, x_2;\; \sigma_{\varphi}, \varphi, \Sigma, m) = \\
  = T_{\varphi^m,\norm}(x_1, x_2;\; \sigma_{\varphi}, \Sigma)
  = \sigma_{\varphi}^{m} \, \partial_{\varphi}^{m} \left( g(x_1, x_2;\; \Sigma) \right),
\end{multline}
where
\begin{itemize}
\item
   $\varphi \in [-\pi, \pi[$ is the preferred orientation of the receptive
   field,
\item
  $\sigma_{\varphi} \in \bbbr_+$ is the amount of spatial smoothing
  in units of the spatial standard deviation,%
\footnote{In the isotropic case, the standard deviation
  $\sigma_{\varphi}$ is related to the variance-based scale parameter $s$ according to
  $s = \sigma_{\varphi}^2$.}
  and which may in a special case be
  defined as
  $\sigma_{\varphi} = \sqrt{e_{\varphi}^T \, \Sigma \, e_{\varphi}}$,
  where $e_{\varphi}$ is the unit vector in the direction of $\varphi$,
\item
  $\partial_{\varphi}^m =
  (\cos \varphi \, \partial_{x_1} + \sin  \varphi \, \partial_{x_2})^m$
  is an $m$th-order directional derivative operator,
  which in a reduced submanifold of receptive field shapes may
  be aligned with the direction of an eigenvector of the spatial
  covariance matrix $\Sigma$, but more generally
  need to be decoupled from that direction to
  allow for full covariance properties,
 \item
   $\Sigma$ is a $2 \times 2$ symmetric positive definite
   covariance matrix, where in a reduced submanifold of the receptive field
   shapes, one may consider one of the eigenvectors of $\Sigma$ as being in
   the direction of $\varphi$; to obtain full ability of the receptive
   fields to span all the degrees of freedom of the geometric image
   transformations, such a criterion about alignment does, however,
   need to be relaxed,
 \item
   $g(x;\; \Sigma)$ is a 2-D affine Gaussian kernel, with its shape
   determined by the spatial covariance matrix $\Sigma$
   \begin{equation}
     \label{eq-2D-aff-gauss}
     g(x;\; \Sigma)
     = \frac{1}{2 \pi \sqrt{\det \Sigma}}
         e^{-x^T \Sigma^{-1} x/2}
    \end{equation}
    for $x = (x_1, x_2)^T \in \bbbr^2$.
\end{itemize}
Concerning time-dependent image data, spatio-temporal receptive fields
corresponding to simple cells
are, in turn, formulated according to
\begin{align}
  \begin{split}
    \label{eq-spat-temp-RF-model-der-norm-caus}
    T_{\text{simple}}(x_1, x_2, t;\; \sigma_{\varphi}, \sigma_t, \varphi, v, \Sigma, m, n) 
  \end{split}\nonumber\\
  \begin{split}
   & = T_{{\varphi}^m, {\bar t}^n,\norm}(x_1, x_2, t;\; \sigma_{\varphi}, \sigma_t, v, \Sigma)
  \end{split}\nonumber\\
  \begin{split}
   &  = \sigma_{\varphi}^{m} \, 
          \sigma_t^{n} \, 
         \partial_{\varphi}^{m} \,\partial_{\bar t}^n 
          \left( g(x_1 - v_1 t, x_2 - v_2 t;\; \Sigma) \,
           h(t;\; \sigma_t) \right),
  \end{split}
\end{align}
where, for the symbols not previously defined in connection
with Equation~(\ref{eq-spat-RF-model}), we have that:
\begin{itemize}
\item
  $\sigma_t$ represents the amount of temporal smoothing in units of
  the temporal standard deviation,%
\footnote{Later, we will also parameterize the temporal scale parameter in
  terms of the variance $\tau = \sigma_t^2$.}
\item
  $v = (v_1, v_2)^T$ represents a local motion vector representing
  the preferred direction of the receptive field,
\item
  $\partial_{\bar t}^n = (\partial_t + v_1 \, \partial_{x_1} + v_2 \, \partial_{x_2})^n$
  represents an $n$th-order velocity-adapted temporal derivative
  operator, and
\item
  $h(t;\; \sigma_t)$ represents a temporal smoothing kernel with temporal
  standard deviation $\sigma_t$.
\end{itemize}
For the case of the temporal domain being non-causal
(meaning that the future relative to an temporal moment can be
accessed, as it can be on pre-recorded video data),
the temporal kernel can be chosen as the 1-D Gaussian kernel
\begin{equation}
  \label{eq-non-caus-temp-gauss}
  h(t;\; \sigma_t) = \frac{1}{\sqrt{2 \pi} \sigma_t} \, e^{-t^2/2\sigma_t^2},
\end{equation}
whereas in the case of the temporal domain being time-causal
(implying the more realistic real-time scenario, where the future cannot be accessed),
the temporal kernel can, based on the requirements of non-creation of
new local extrema over the purely temporal domain in combination of
a choice of the time constants of the resulting truncated exponential
kernels to correspond to temporal scale covariance,
be determined as the time-causal limit kernel
(see Lindeberg \citeyear{Lin21-Heliyon} Appendix~B for a derivation)
defined according to
(Lindeberg \citeyear{Lin16-JMIV} Section~5,
Lindeberg \citeyear{Lin23-BICY} Section~3)
\begin{equation}
  \label{eq-time-caus-lim-kern}
  h(t;\; \sigma_t) = \Psi(t;\; \sigma_t, c),
\end{equation}
characterized by having a Fourier transform of the form
\begin{equation}
  \label{eq-FT-comp-kern-log-distr-limit}
     \hat{\Psi}(\omega;\; \sigma_t, c) 
     = \prod_{k=1}^{\infty} \frac{1}{1 + i \, c^{-k} \sqrt{c^2-1} \, \sigma_t \, \omega}.
\end{equation}
This form of the temporal smoothing function corresponds to using
an infinite set of first-order integrators that are coupled in
cascade
\begin{equation}
  \label{eq-def-time-caus-lim-kern-inf-conv}
  \Psi(\cdot;\; \sigma_t, c) = *_{k=1}^{\infty} h_{\exp}(\cdot;\; \mu_k),
\end{equation}
with the time constants
\begin{equation}
  \mu_k = c^{-k} \sqrt{c^2-1} \, \sigma_t
\end{equation}
chosen so as to specifically enable temporal
scale covariance. The distribution parameter $c > 1$, often chosen
as $c = \sqrt{2}$ or $c = 2$, in this temporal
smoothing function reflects the ratio between adjacent discrete
temporal scale levels in the corresponding temporal
scale-space model. For practical implementation purposes, the infinite
number of first-order integrators can often be truncated after the 4
to 8 slowest time constants (Lindeberg \citeyear{Lin26-FrontSignProc}).

In Lindeberg (\citeyear{Lin21-Heliyon}), it was demonstrated
that idealized receptive field models of these types do
rather well model the qualitative shape of biological
simple cells, as obtained from neurophysiological measurements
by DeAngelis {\em et al.\/}\
(\citeyear{DeAngOhzFre95-TINS,deAngAnz04-VisNeuroSci}),
Conway and Livingstone (\citeyear{ConLiv06-JNeurSci}) and
Johnson {\em et al.\/}\ (\citeyear{JohHawSha08-JNeuroSci}).
See Figures~12--18 in Lindeberg (\citeyear{Lin21-Heliyon})
for comparisons between biological receptive fields and idealized
models thereof, based on the generalized Gaussian derivative model for
visual receptive fields.

\begin{figure*}[hbtp]
  \begin{subfigure}{0.99\textwidth}
    \begin{center}
      \begin{tabular}{cccccc}
        $\bar{\sigma} = 1/\sqrt{2}$
        & $\bar{\sigma} = 1$
        & $\bar{\sigma} = \sqrt{2}$
        & $\bar{\sigma} = 2$
        & $\bar{\sigma} = 2\sqrt{2}$
        & $\bar{\sigma} = 4$ \\
        \includegraphics[width=0.14\textwidth]{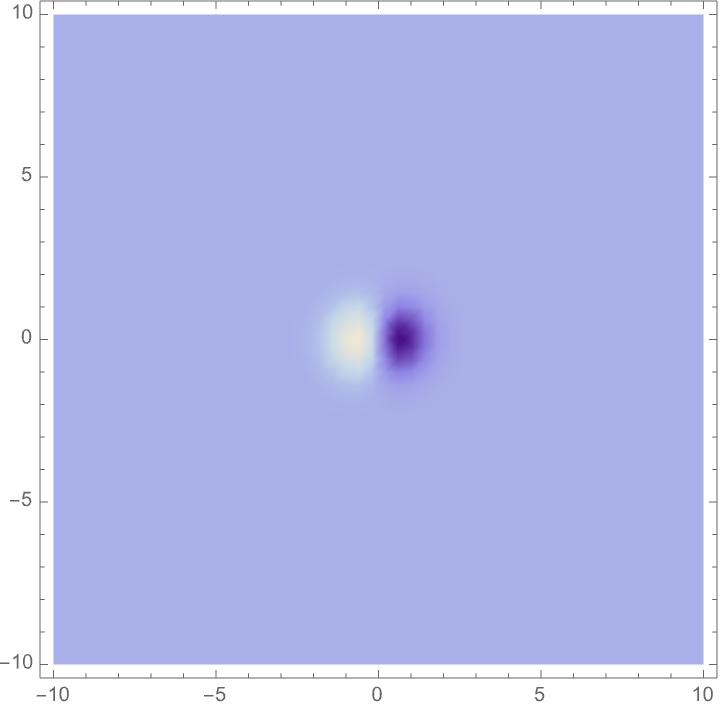}
        & \includegraphics[width=0.14\textwidth]{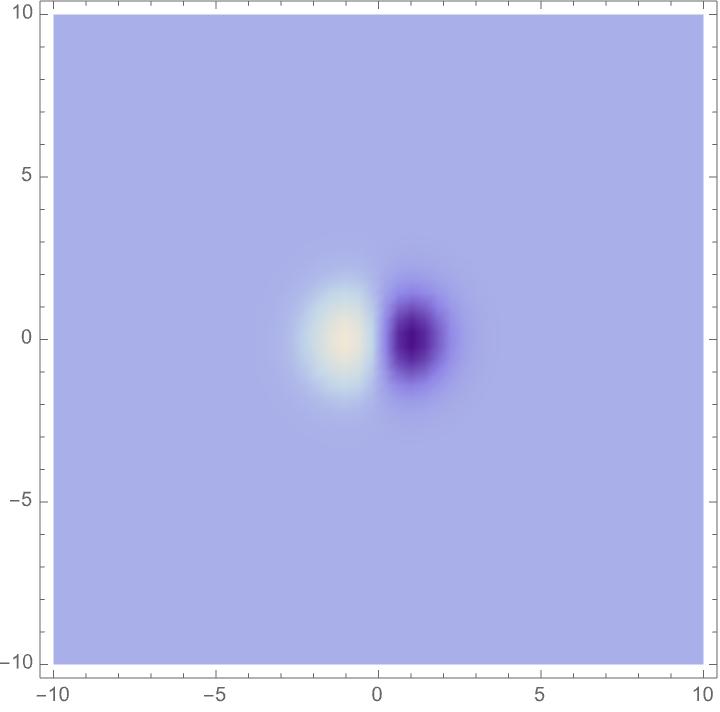}
        & \includegraphics[width=0.14\textwidth]{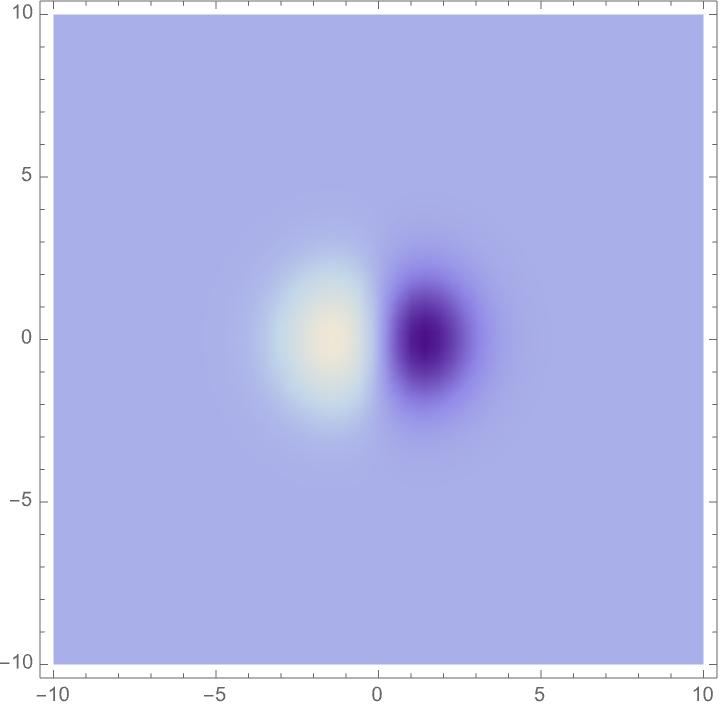}
        & \includegraphics[width=0.14\textwidth]{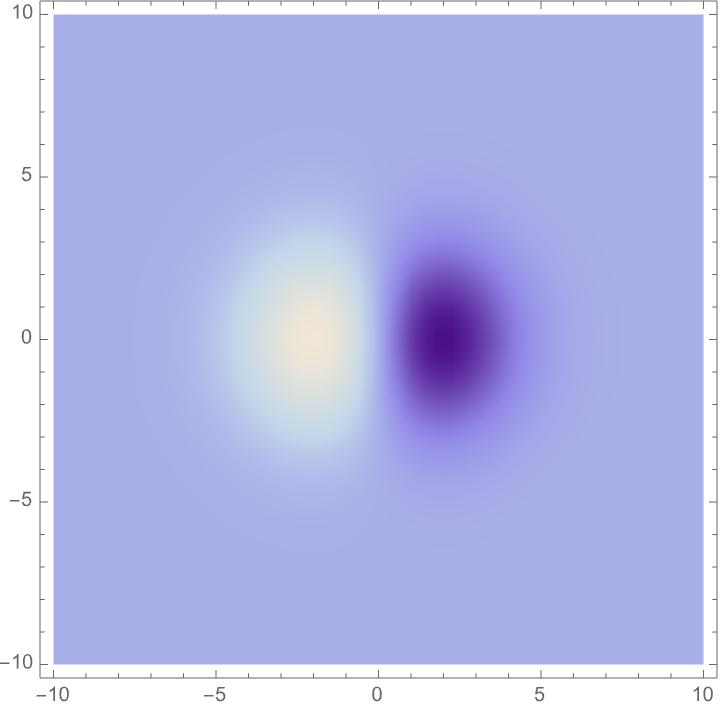}
        & \includegraphics[width=0.14\textwidth]{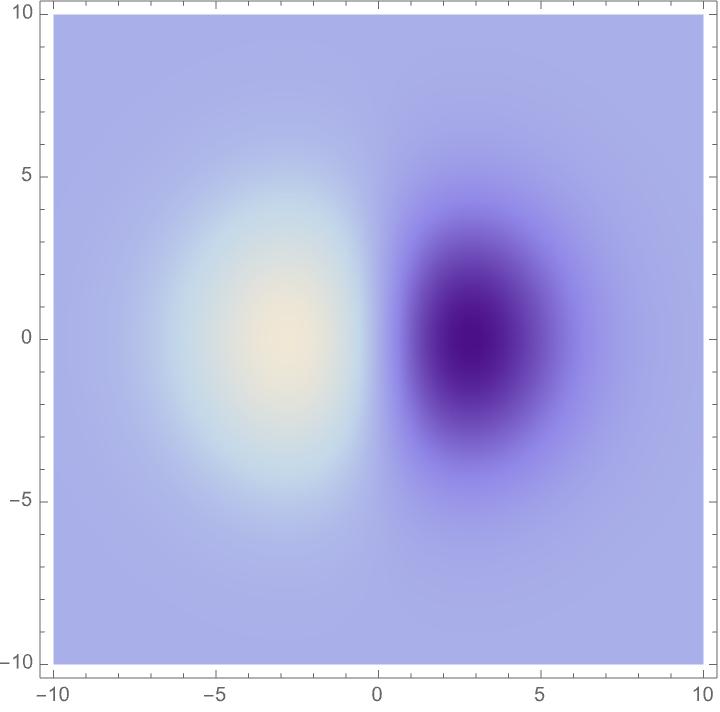}
        & \includegraphics[width=0.14\textwidth]{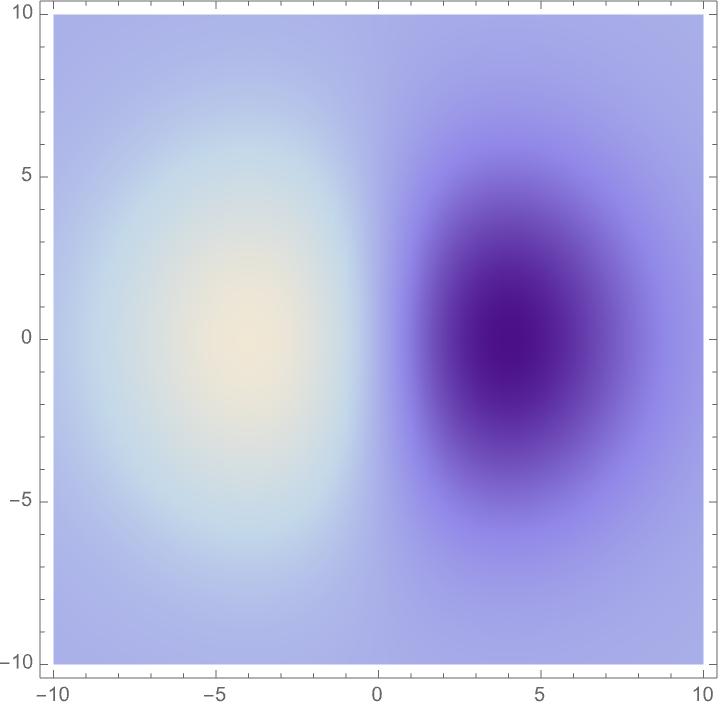} \\
      \end{tabular}
    \end{center}
    \caption{{\em Variability in the size \/} of first-order affine Gaussian derivative
      receptive fields (for $\sigma_1 = \sigma_2$ and
      image orientation $\varphi = 0$),
      with the overall size $\bar{\sigma} =
      \sqrt{\sigma_1 \sigma_2}$ increasing from $1/\sqrt{2}$ to $4$
      according to a logarithmic distribution, from left to right.
      ({\bf Horizontal axes:} image coordinate $x_1 \in [-10, 10]$.
      {\bf Vertical axes:} image coordinate $x_2 \in [-10, 10]$.)}
    \label{fig-size-variability}
  \end{subfigure}

  \bigskip
  
  \begin{subfigure}{0.99\textwidth}
    \begin{center}
      \begin{tabular}{cccccc}
        $\varphi = 0$
        & $\varphi= \frac{\pi}{6}$
        & $\varphi= \frac{\pi}{3}$
        & $\varphi= \frac{\pi}{2}$
        & $\varphi= \frac{2\pi}{3}$
        & $\varphi= \frac{5\pi}{6}$ \\
        \includegraphics[width=0.14\textwidth]{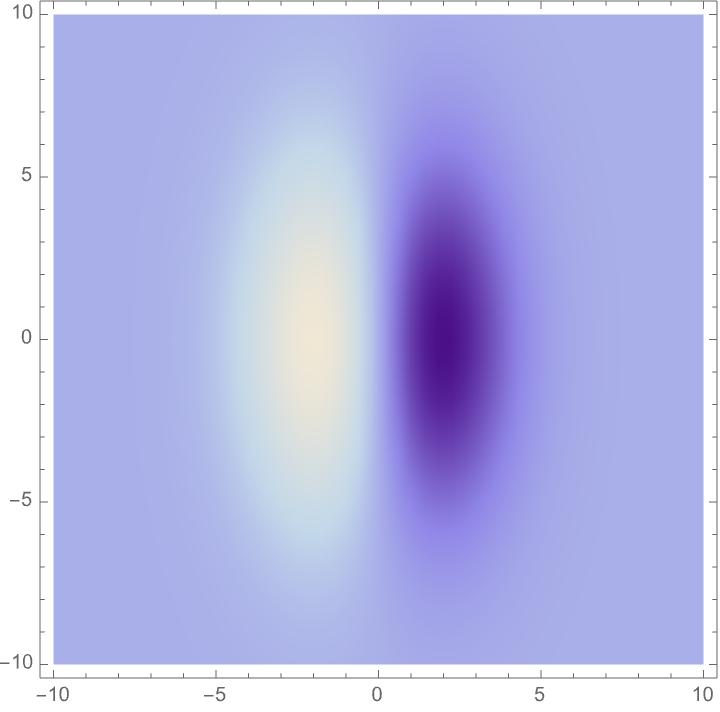}
        & \includegraphics[width=0.14\textwidth]{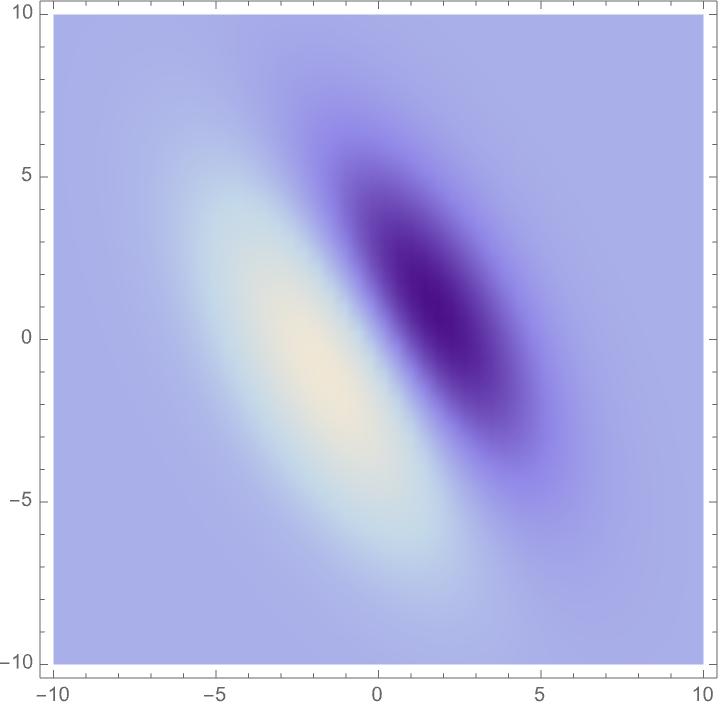}
        & \includegraphics[width=0.14\textwidth]{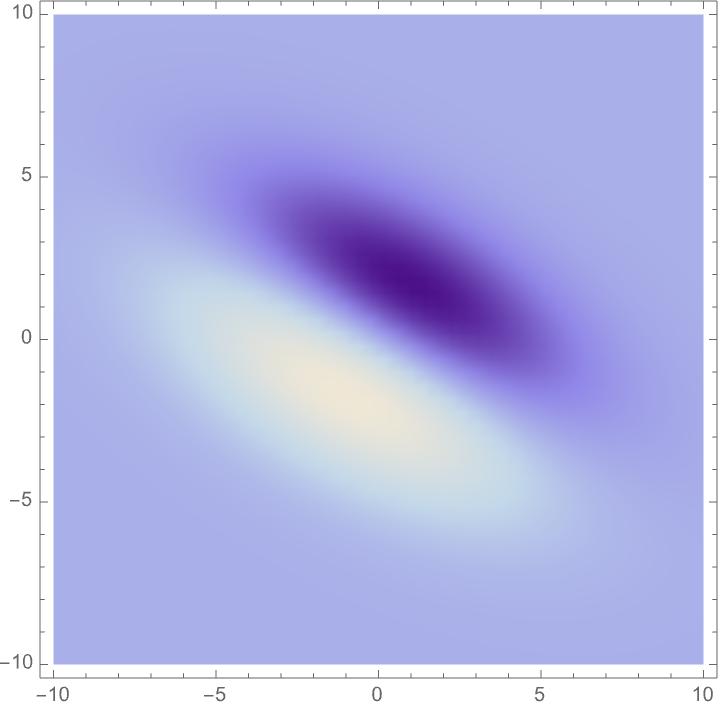}
        & \includegraphics[width=0.14\textwidth]{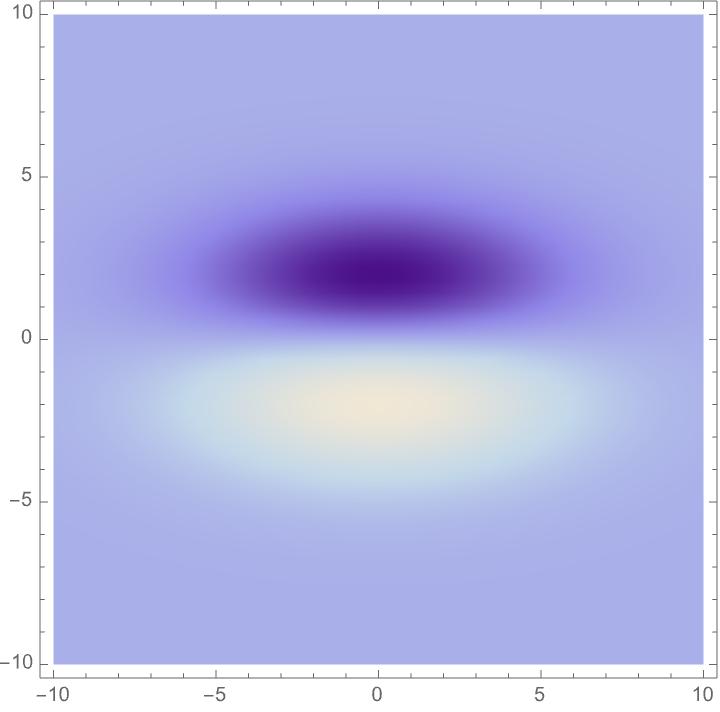}
        & \includegraphics[width=0.14\textwidth]{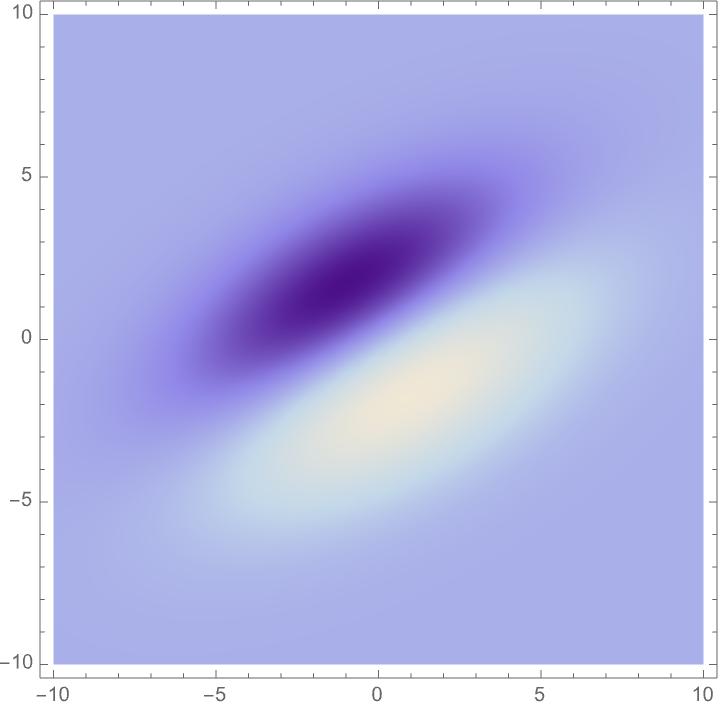}
        & \includegraphics[width=0.14\textwidth]{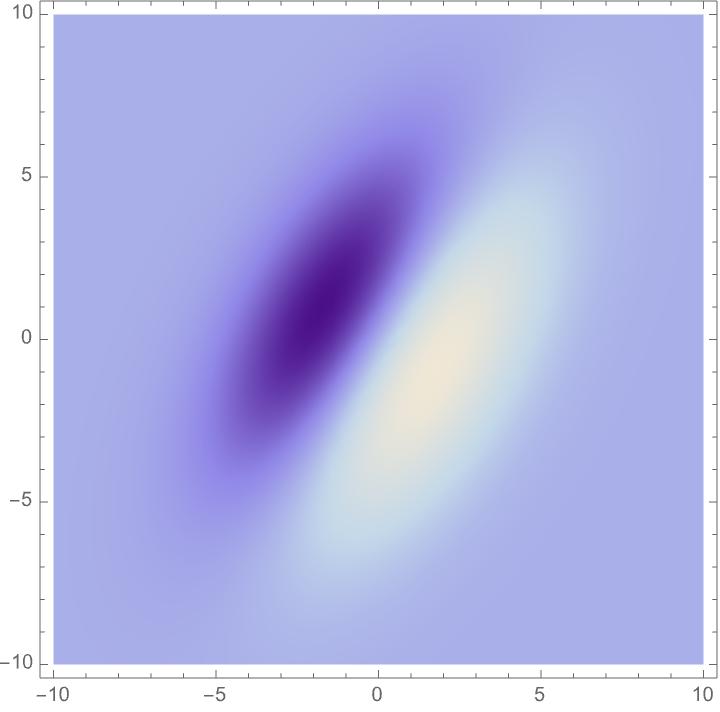} \\
    \end{tabular}
    \end{center}
    \caption{{\em Variability in the orientation\/} of first-order
      affine Gaussian derivative
      receptive fields (for $\sigma_1 = 2$ and $\sigma_2 = 4$),
      with the orientation angle $\varphi$ increasing from left to right.
     ({\bf Horizontal axes:} image coordinate $x_1 \in [-10, 10]$.
     {\bf Vertical axes:} image coordinate $x_2 \in [-10, 10]$.)}
    \label{fig-ori-variability}
  \end{subfigure}

  \bigskip

  \begin{subfigure}{0.99\textwidth}
    \begin{center}
      \begin{tabular}{cccccc}
        $\kappa= 1$
        & $\kappa= \sqrt{2}$
        & $\kappa= 2$
        & $\kappa= 2\sqrt{2}$
        & $\kappa= 4$        
        & $\kappa= 4\sqrt{2}$ \\
        \includegraphics[width=0.14\textwidth]{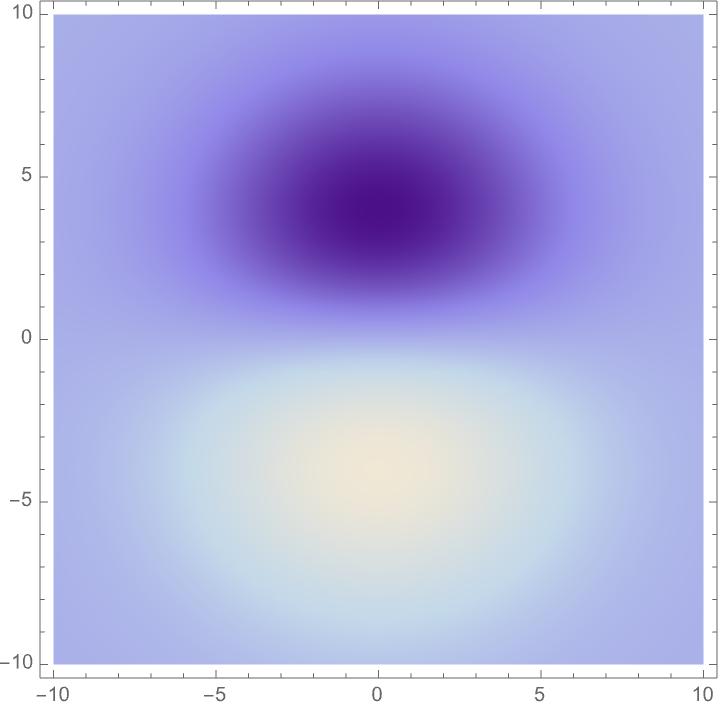}
        & \includegraphics[width=0.14\textwidth]{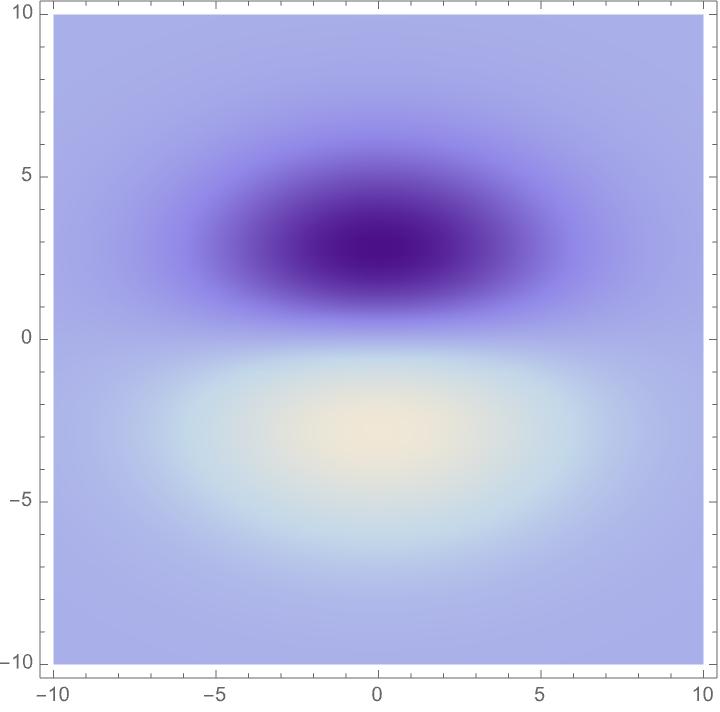}
        & \includegraphics[width=0.14\textwidth]{affgauss-1dirder-sigma-2-4-phi-piover2-mono.jpg}
        & \includegraphics[width=0.14\textwidth]{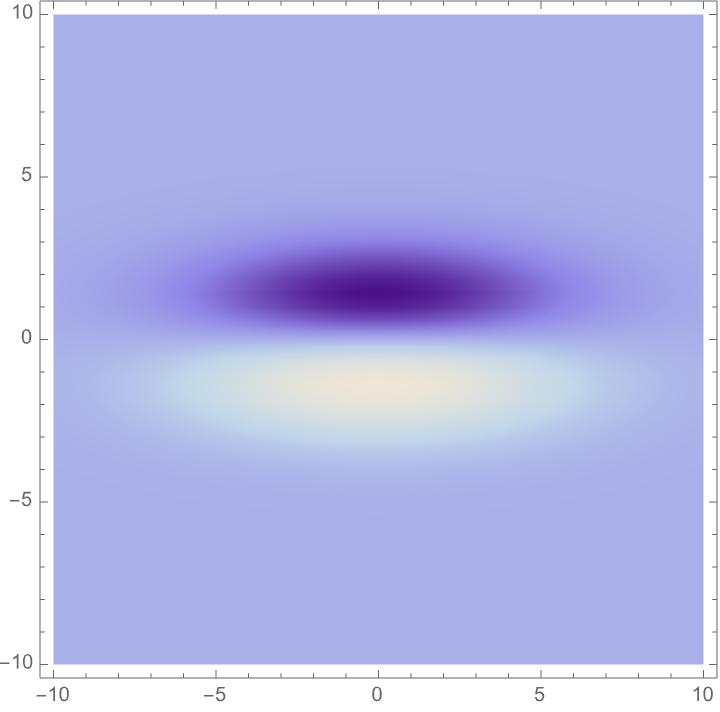}
        & \includegraphics[width=0.14\textwidth]{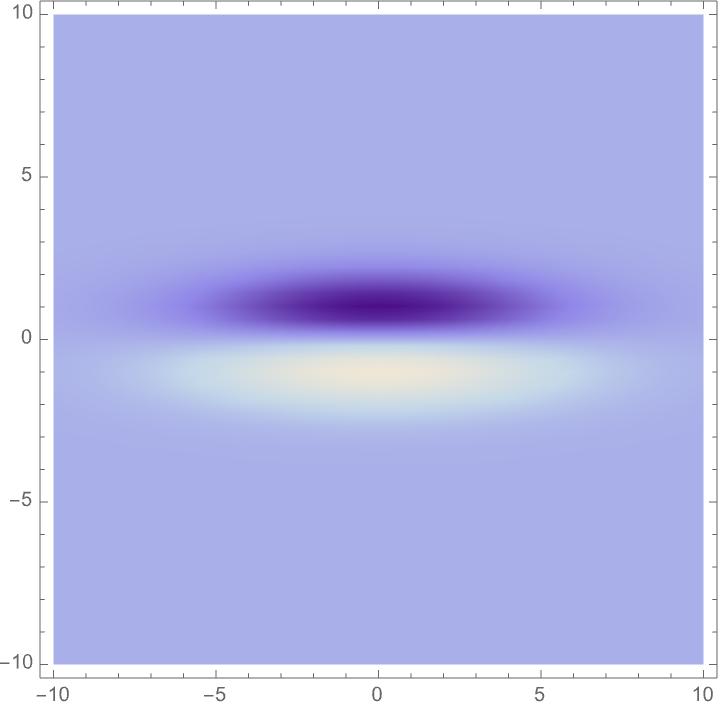}
        & \includegraphics[width=0.14\textwidth]{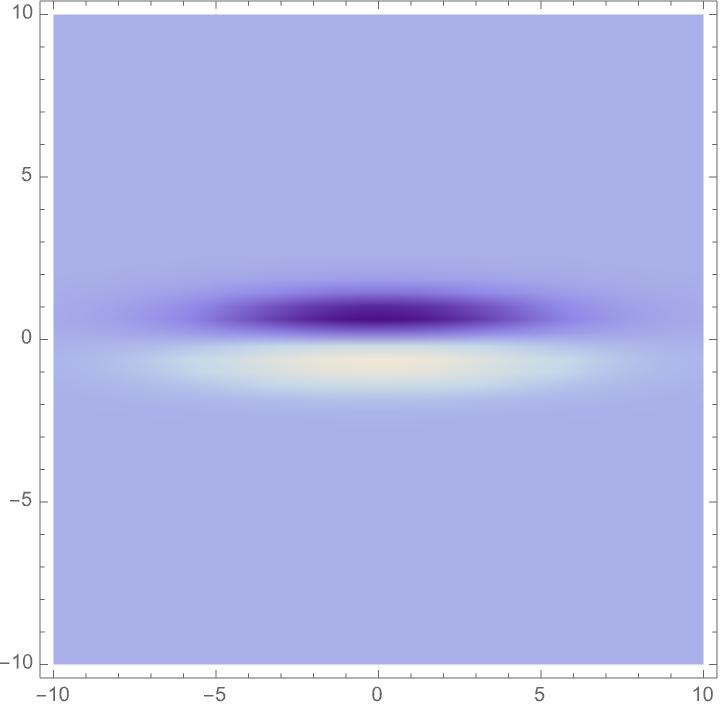} \\ 
      \end{tabular}
    \end{center}
    \caption{{\em Variability in the degree of elongation\/} of
      first-order affine Gaussian derivative
      receptive fields (for image orientation $\varphi = \frac{\pi}{2}$),
      with the eccentricity, for these Cartesian-aligned kernels given by
      $\kappa = \sigma_1/\sigma_2$, increasing from $1$ to $4\sqrt{2}$
      from left to right, with $\sigma_1$ kept constant.
      ({\bf Horizontal axes:} image coordinate $x_1 \in [-10, 10]$.
      {\bf Vertical axes:} image coordinate $x_2 \in [-10, 10]$.)}
    \label{fig-ecc-variability}
  \end{subfigure}

  \bigskip
  
  \begin{subfigure}{0.99\textwidth}
    \begin{center}
      \begin{tabular}{cccccc}
        & $v = -1$ & $v = -1/2$ & $v = 0$ & $v = 1/2$ & $v = 1$
        \\
        $\scriptsize{\sigma_x = 2, \sigma_t = 2}$
        & \includegraphics[width=0.14\textwidth]{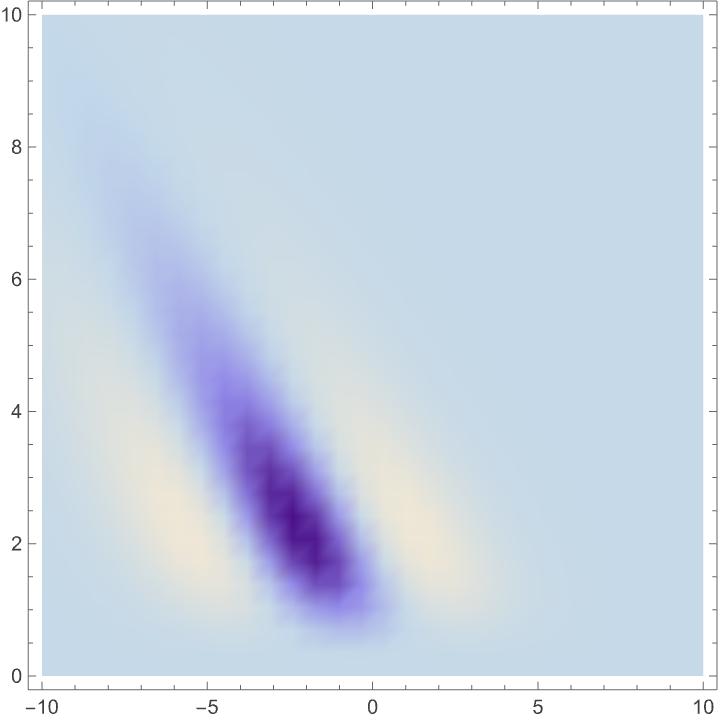}
                   & \includegraphics[width=0.14\textwidth]{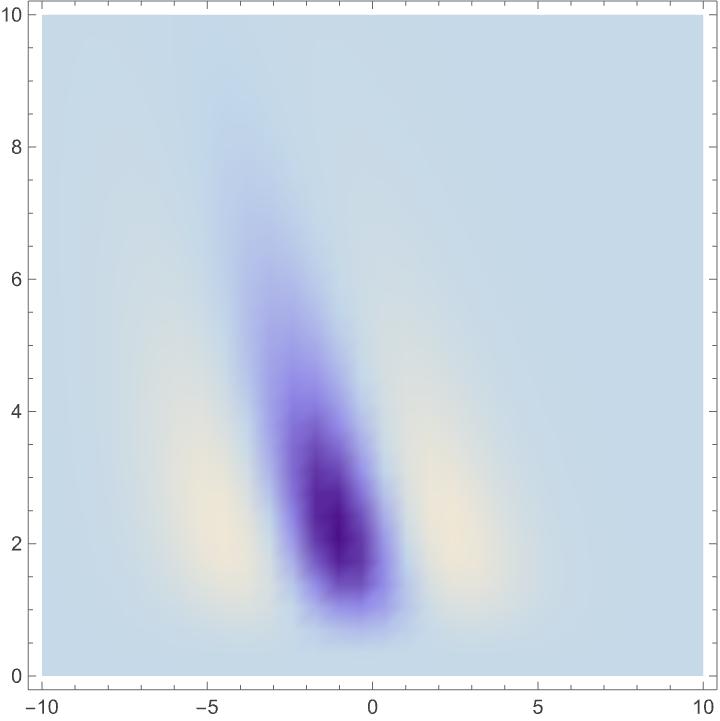}
                                & \includegraphics[width=0.14\textwidth]{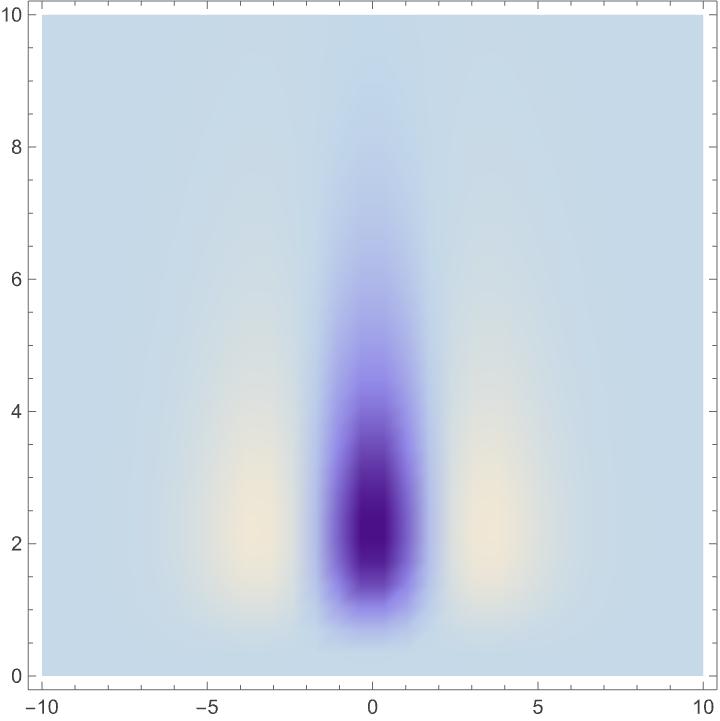}
                                          & \includegraphics[width=0.14\textwidth]{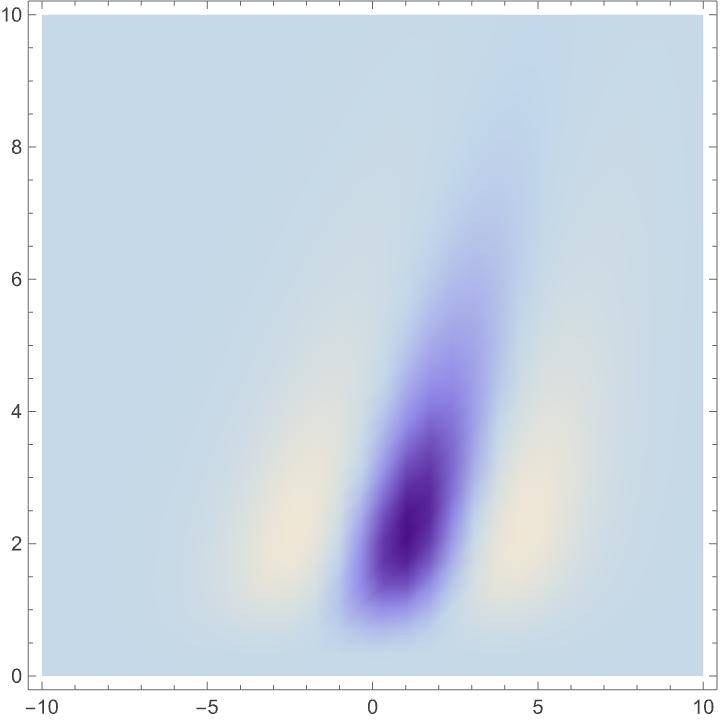}
                                                      & \includegraphics[width=0.14\textwidth]{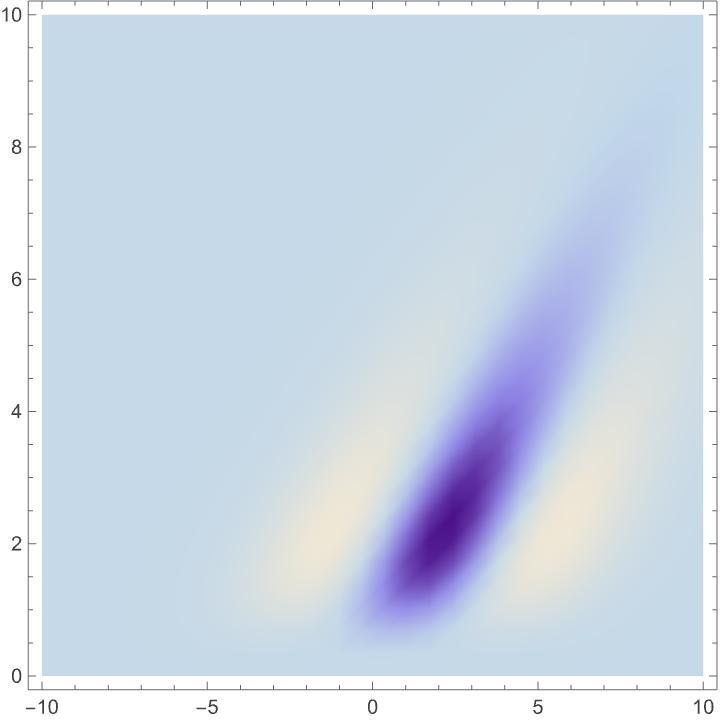}                                          
        \\
        $\scriptsize{\sigma_x = 2, \sigma_t = 1}$
        & \includegraphics[width=0.14\textwidth]{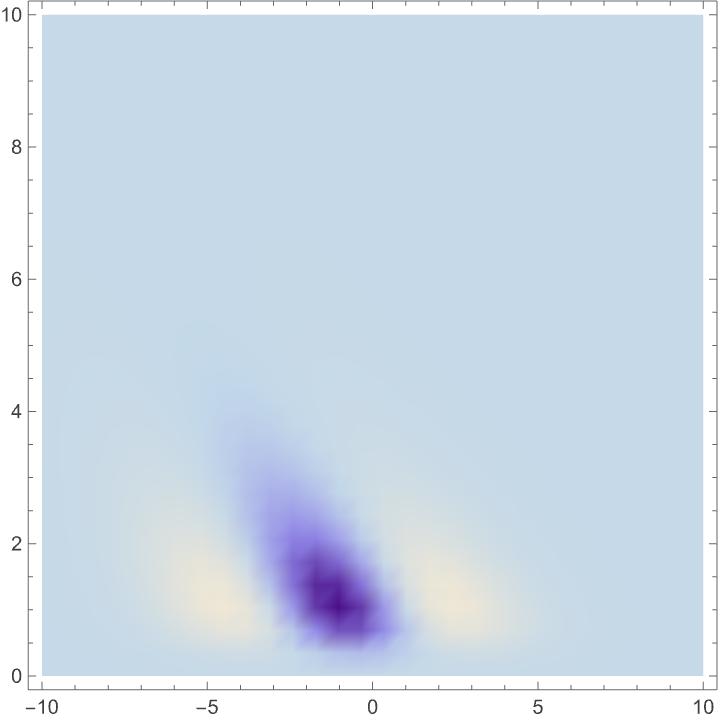}
                   & \includegraphics[width=0.14\textwidth]{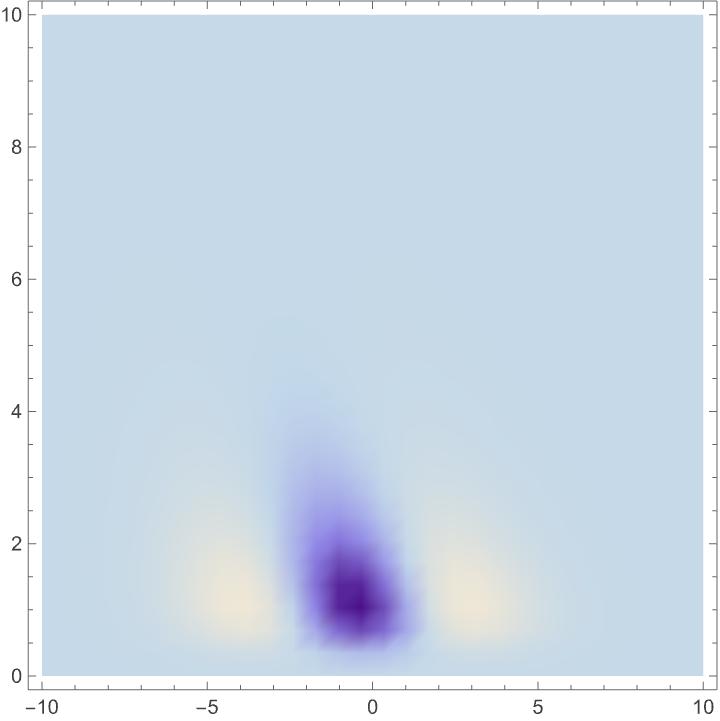}
                                & \includegraphics[width=0.14\textwidth]{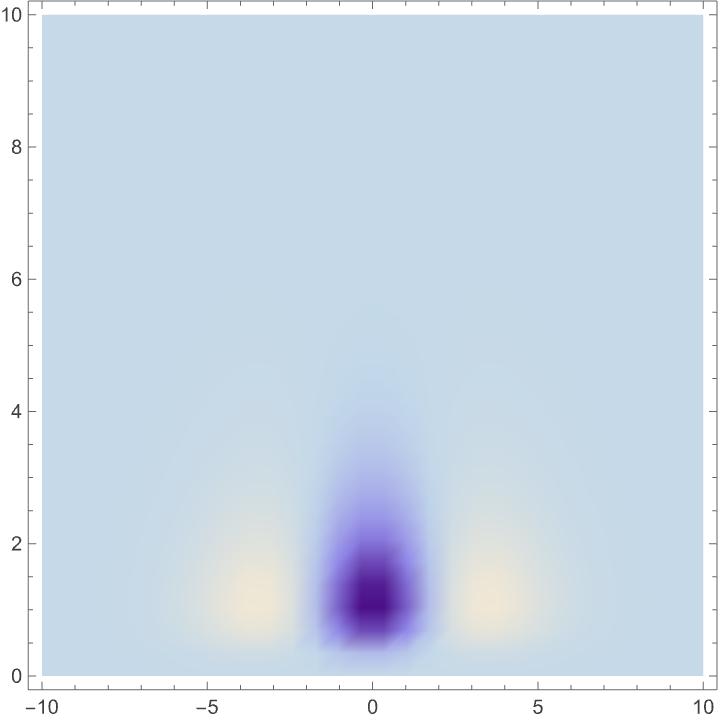}
                                          & \includegraphics[width=0.14\textwidth]{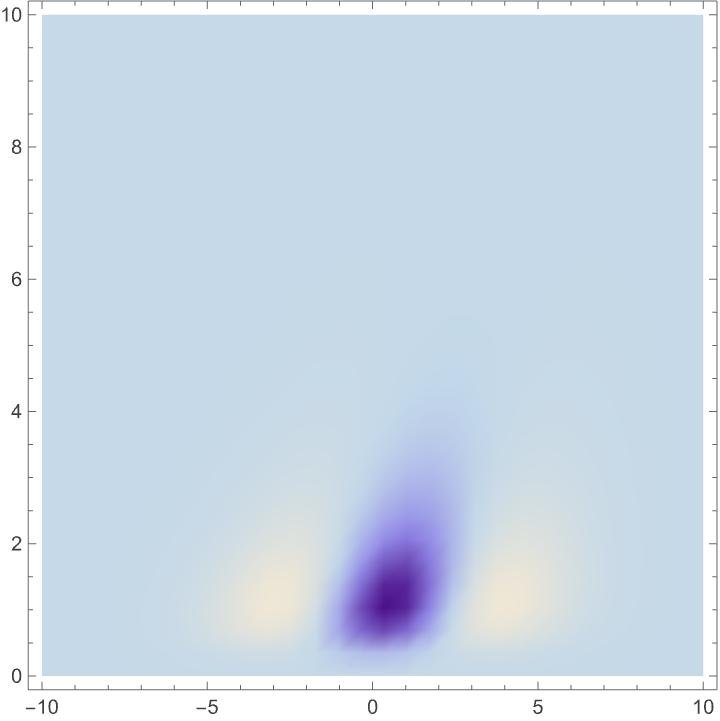}
                                                      & \includegraphics[width=0.14\textwidth]{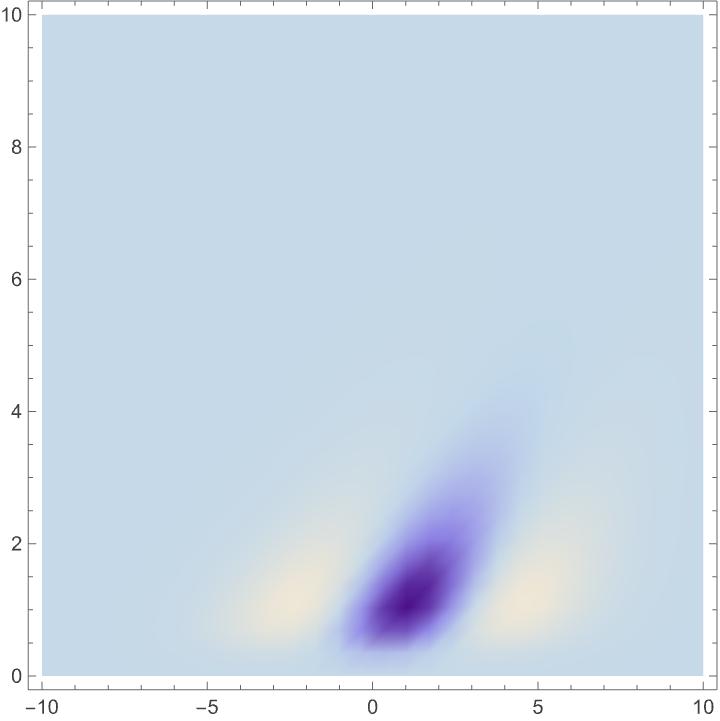}                                          
        \\
      \end{tabular}
    \end{center}
    \caption{{\em Variabilities along Galilean transformations and over temporal scales\/}
      for time-causal joint spatio-temporal receptive fields over
      a 1+1D spatio-temporal domain in terms of the second-order
      spatial derivative of the form $T_{xx,\text{norm}}(x, t;\; s, \tau, v)$
      according to (\ref{eq-spat-temp-RF-model-der-norm-caus}) of the product of a
      velocity-adapted 1-D Gaussian kernel
      over the spatial domain and the time-causal limit kernel over the
      temporal domain according to (\ref{eq-time-caus-lim-kern}).
      ({\bf Horizontal axes:} spatial image coordinate $x \in [-10, 10]$.
      {\bf Vertical axes:} temporal variable $t \in [0, 8]$.)}
    \label{fig-timecaus-strfs}
  \end{subfigure}

  \caption{Variabilities in the shape of spatial and spatio-temporal
    receptive fields {\bf (a)}~over spatial scale $\sigma_{\varphi}$,
    {\bf (b)}~over image orientations $\varphi$,
    {\bf (c)}~over the degree of elongation, and
    {\bf (d)}~over image velocity $|v|$ and temporal scale $\sigma_t$.}
  \label{fig-rf-variabilities}
\end{figure*}

\section{Variabilities in receptive field shapes as implied by
  covariance properties under geometric image transformations}

A characteristic property of the generalized Gaussian derivative
theory for visual receptive fields, is that the shapes of the
receptive fields are parameterized in terms of a set of multiple 
parameters.

\subsection{Degrees of freedom in the generalized Gaussian derivative
  model}

Using the following measure of the degree of elongation
of the spatial covariance matrix $\Sigma$ 
\begin{equation}
  \kappa = \frac{\sigma_{\orth}}{\sigma_{\varphi}},
\end{equation}
where $\sigma_{\orth} = \sqrt{e_{\orth}^T \, \Sigma \, e_{\orth}}$
represents the amount of smoothing in the direction $e_{\orth}$
perpendicular to the preferred direction $e_{\varphi}$,
the model comprises the following total number of 7~degrees of freedom:
\begin{itemize}
\item
  the spatial scale parameter $\sigma_{\varphi}$,
\item
  the preferred orientation $\varphi$,
\item
  the degree of elongation $\kappa$,
\item
  a fourth degree of freedom corresponding to the angle between the
  preferred direction $\varphi$ and the direction of the main
  eigenvector of the spatial covariance matrix $\Sigma$,
\item
  the preferred direction $\arg v = \operatorname{arctan2}(v_1, v_2)$
  in joint space-time,
\item
  the preferred speed $|v| = \sqrt{v_1^2 + v_2^2}$, and
\item
  the temporal scale $\sigma_t$,
\end{itemize}
see Figure~\ref{fig-rf-variabilities} for visualizations of a number
of these variabilities in receptive field shapes.

According to the most idealized version of the theory, there ought to
be variabilities in the shapes of the receptive fields along all these
degrees of freedom (Lindeberg \citeyear{Lin26-JCompNeuroSci}).

According to the theory for cascade smoothing
properties over spatial scales in Lindeberg (\citeyear{Lin26-JMIV}),
a vision system could,
however, also choose to only represent the finest scales explicitly in
the sensorium corresponding to the primary visual cortex.
Equivalent representations at coarser scales could then instead be
computed in an indirect manner at later stages in the visual
hierarchy.

\subsection{Prediction about variabilities in the shapes of the
  receptive fields based on covariance properties under geometric
  image transformations}

The motivation behind predicting such variabilities in the shapes of
the receptive fields is to enable a matching between receptive field
responses computed under different viewing conditions.
In general, the response of any specific receptive field will be
substantially affected by viewing variations, due to
(i)~different distances between the object/event and the observer,
(ii)~different viewing directions in relation to the local normal
directions of the object(s),
(iii)~different relative motions between the object/event and the observer,
and (iv)~different speeds by which a spatio-temporal event may occur,
Hence, under changes in the viewing conditions, it would not
be possible to in any direct way match the receptive field responses
from the same receptive field before and after the transformation.

\begin{figure*}[hbtp]
  \begin{center}
    \begin{subfigure}{0.98\textwidth}
      \begin{center}
        \begin{tabular}{cc}
          {\em\small Non-covariant receptive fields\/} & {\em\small Covariant receptive fields\/} \\
          \includegraphics[width=0.40\textwidth]{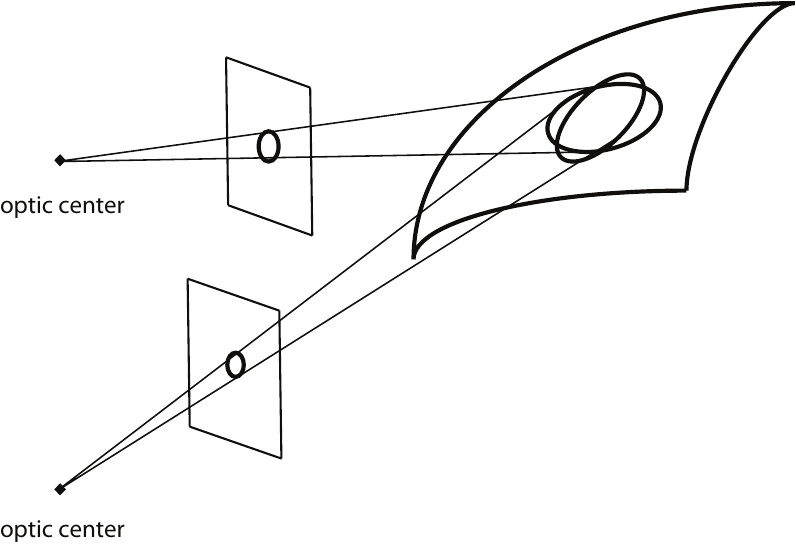}
          & \includegraphics[width=0.40\textwidth]{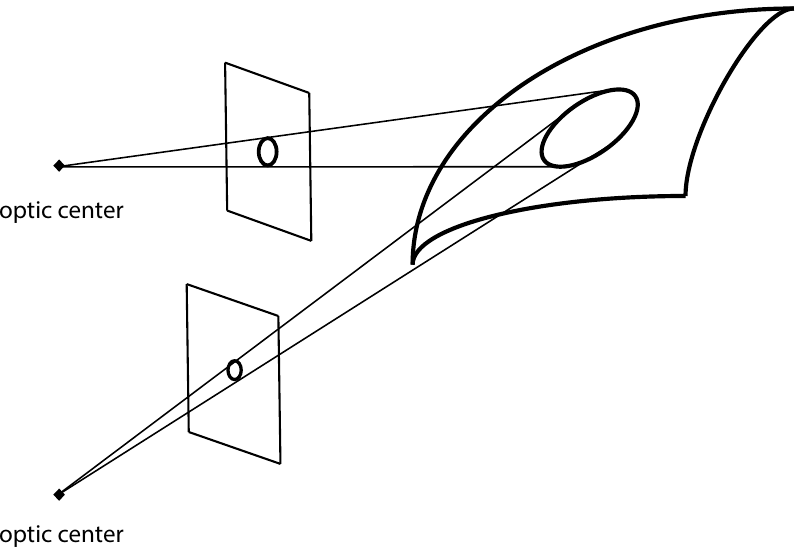} \\
        \end{tabular}
      \end{center}
      \caption{Illustration of the importance of covariance properties of
    the receptive fields when computing receptive field responses of a
    scene under different viewing conditions.
    (left) If the receptive field family is not covariant with respect
    to the appropriate class of geometric image transformations, then the
    backprojections of the receptive fields onto the tangent plane of
    an observed local surface patch will, in general, be different in the tangent
    plane of the surface. Thereby, if the receptive field responses
    are to be used for, to example, for computing local shape
    properties of the surface patch, then those shape estimates may be
    strongly biased because of effects of that mismatch between the
    backprojected receptive fields.
    (right) If the receptive field family is covariant with respect to
    the relevant class of geometric image transformations, then it is,
    on the other hand, possible to match the parameters of the
    receptive fields over the two image domains in such a way that the
    backprojected receptive fields do to first order of approximation
    coincide in the tangent plane of the surface. Thereby, the source
    to bias caused by a mismatch of the backprojected receptive fields
    can be substantially reduced, which enables to computation of more
    accurate estimates of the local surface shape.
    While this example concerns spatial receptive fields corresponding
    to two views with different viewing directions relative to a
    static scene, corresponding effects regarding the backprojections
    of the receptive fields will occur also when computing
    spatio-temporal receptive field responses for dynamic scenes.
    (Figure reproduced from Lindeberg
    \citeyear{Lin23-FrontCompNeuroSci} with permission (Open Access).)} 
      \label{fig-backproj-rfs-rot-symm-vs-aff}
    \end{subfigure}
  \end{center}
  \bigskip
  \begin{center}
    \begin{subfigure}{0.49\textwidth}
      \vspace{-40mm}
      \includegraphics[width=0.99\textwidth]{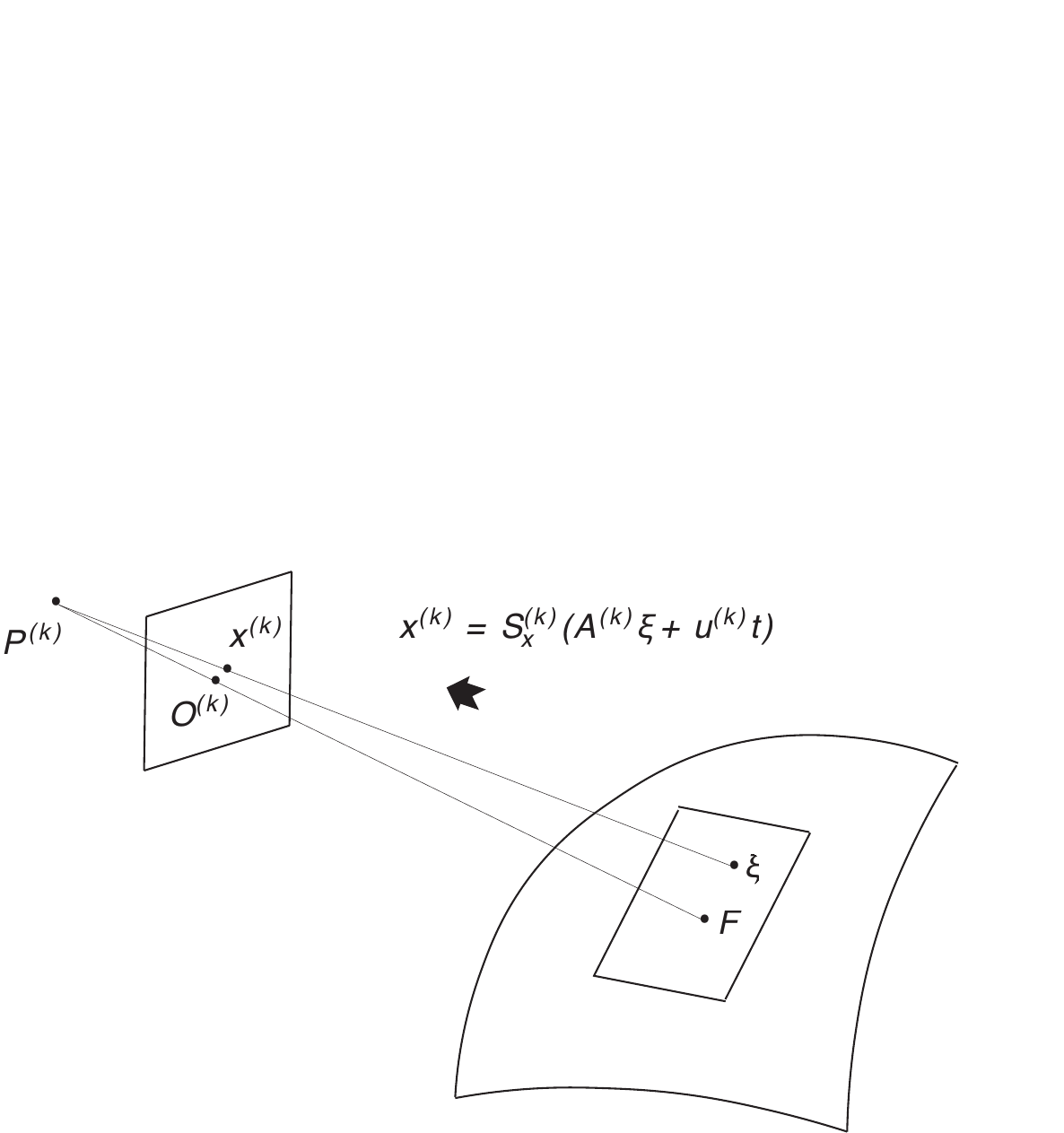} 
      \caption{Illustration of the geometry underlying the composed
        locally linearized projection model in Equations~(\ref{eq-x-transf}) and
        (\ref{eq-t-transf}) for a single monocular view.
        Here, a local, possibly moving, surface patch is projected to an
        arbitrary view indexed by $k$ in a multi-view locally linearized
        projection model, with
        the fixation point $F$ on the surface mapped to the origin
        $O^{(k)} = 0$ in the image plane for the observer with the
        optic center $P^{(k)}$. Then, any point in the
        tangent plane to the surface at the fixation point, as
        parameterized by the local coordinates $\xi$ in a coordinate
        frame attached to the tangent plane of the surface with $\xi = 0$
        at the fixation point $F$, is by the
        local linearization mapped to the image point $x^{(k)}$.
        (Figure reproduced from Lindeberg (\citeyear{Lin25-JMIV}) with
        permission (OpenAccess).)}
      \label{fig-singlegeom}
    \end{subfigure}
    \hfill
    \begin{subfigure}{0.49\textwidth}
      \vspace{-25mm}
      \includegraphics[width=0.99\textwidth]{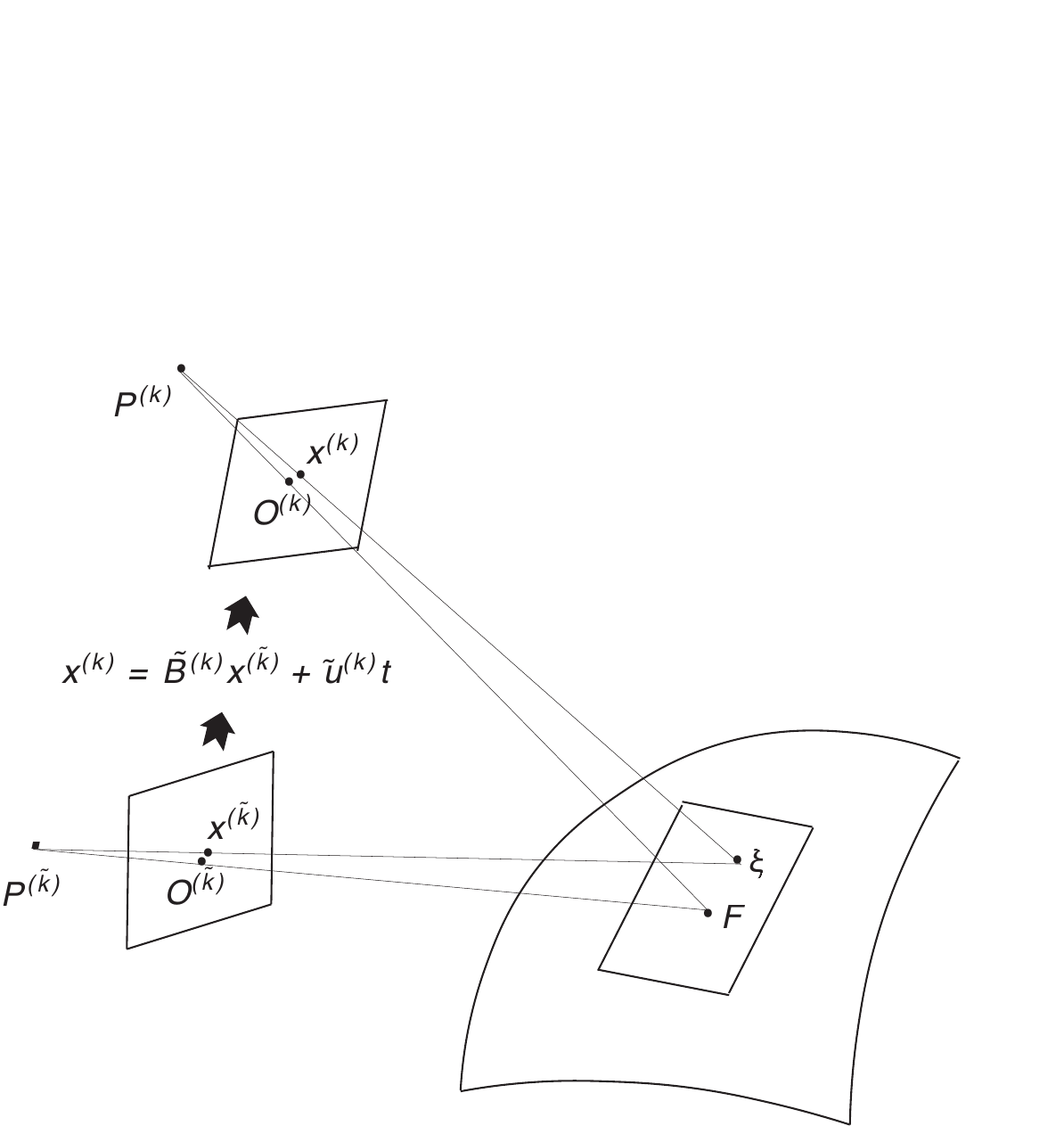} 
      \caption{Illustration of the underlying geometric situtation for
        the locally linearized transformations between pairwise views of
        the same, possibly moving, local surface patch
        according to Equation~(\ref{eq-sc-aff-vel-transf-alt-obs-model}).
        Here, the view
        indexed by $\tilde{k}$ constitutes the reference view and the
        view indexed by $k$ constitutes an arbitrary view.
        By a connection of the point $\xi$ being the same in two instances
        of Figure~\ref{fig-singlegeom}, we can from the parameters $S_t$,
        $A$ and $u$ of the monocular mappings for optical centers based on
        the indices $k$ and $\tilde{k}$ establish a relationship between
        the matching image points $x^{(k)}$ and $x^{(\tilde{k})}$ in these
        two views. (Figure reproduced from Lindeberg (\citeyear{Lin25-JMIV}) with
        permission (OpenAccess).)}
      \label{fig-pairgeom}   
    \end{subfigure}
  \end{center}
  \caption{Illustrations of {\bf (a)} the importance of covariance to be
    able to match receptive field responses between different views
    and {\bf (b)--(c)} the geometric models for the monocular
    and binocular/multi-view cases.}
\end{figure*}

According to the generalized Gaussian derivative theory,
the receptive fields are covariant under first-order local linearizations of the
geometric image transformations. They transform
in a predictable way under (i)~spatial scaling transformations,
(ii)~spatial affine transformations,
(iii)~Galilean transformations, and
(iv)~temporal scaling transformations
(Lindeberg \citeyear{Lin23-FrontCompNeuroSci,Lin25-JMIV}).
This implies that, if the vision system explicitly
represents a rich family of receptive field
responses for different values of the shape parameters,
then it will be possible to match the receptive field
response for one member in the first view
to the receptive field response for another member
in the second view
(Lindeberg \citeyear{Lin26-JCompNeuroSci}),
see Figure~\ref{fig-backproj-rfs-rot-symm-vs-aff}
for an illustration.

\subsection{Transformation properties of receptive field parameters
  under composed geometric image transformations}

Formally, considering the compositions of these geometric transformations
in the following way when observing dynamic scenes with either
monocular or binocular/multi-view locally linearized camera models
(Lindeberg \citeyear{Lin25-JMIV} Equations~(222)--(223)):
\begin{itemize}
\item
  {\bf Monocular projection model:}
  \begin{align}
    \begin{split}
      \label{eq-x-transf}
      x' = S_x \, (A \,  x + u \, t),
    \end{split}\\
    \begin{split}
      \label{eq-t-transf}
      t' = S_t \, t,
    \end{split}
  \end{align}
  where
  \begin{itemize}
  \item
    the $2 \times 2$ affine transformation matrix $A$ models the
    orthographic projection of surface patterns from the tangent
    plane of a local surface patch to a plane, parallel with the
    image plane of the observer,
  \item
    the velocity vector $u = (u_1, u_2)^T \in \bbbr^2$ models
    the projection of the 3-D motion vector
    $U = (U_1, U_2, U_3)^T$ of local surface patterns onto a plane,
    parallel to the image plane, by local orthographic projection,
  \item
    the spatial scaling factor $S_x \in \bbbr_+$ models the perspective
    scaling factor proportional to the inverse depth $Z$, which will
    then affect both the projection of a spatial surface pattern and the
    magnitude of the perceived motion in the image plane, and
  \item
    the temporal scaling factor $S_t \in \bbbr_+$ models the variability of
    similar spatio-temporal events, that may occur either faster or
    slower, when observing different instances of a similar event at
    different occasions,
  \end{itemize}
  see Figure~\ref{fig-singlegeom} for an illustration.
\end{itemize}
\begin{itemize}
\item
  {\bf Binocular and multi-view projection model:} Given multiple
  views, as indexed by $k$ and $\tilde{k}$ of the same object, with
  \begin{equation}
    \tilde{B}^{(k)}
    = \frac{S_x^{(k)}}{S_x^{(\tilde{k})}} \, A^{(k)} (A^{(\tilde{k})})^{-1}
  \end{equation}
  and
  \begin{equation}
    \tilde{u}^{(k)} = S_x^{(k)} \, A^{(k)} \left( u^{(k)} - u^{(\tilde{k})} \right),
  \end{equation}
  we have that the relations between corresponding image points $x^{(k)}$
  and $x^{(\tilde{k})}$ in these views can be
  expressed as (Lindeberg \citeyear{Lin25-JMIV} Equation~(299))
  \begin{equation}
    \label{eq-sc-aff-vel-transf-alt-obs-model}
    x^{(k)} = \tilde{B}^{(k)} \, x^{(\tilde{k})} + \tilde{u}^{(k)} \, t,
  \end{equation}
  where
  \begin{itemize}
  \item
    $x^{(k)} \in \bbbr^2$ is the locally linearized projection of the physical point
    on the surface pattern in the view from the observer with index $k$
    at time $t$,
  \item
    $x^{(\tilde{k})} \in \bbbr^2$ is the locally linearized projection of the physical point
    on the surface pattern in the view from the observer with index
    $\tilde{k}$ at time $t$,
  \item
    $\tilde{B}^{(k)}$ is a non-singular $2 \times 2$ affine projection matrix
    for the observer with index $k$ in relation to an observation from 
    a reference view with index $\tilde{k}$, and
  \item
    $\tilde{u}^{(k)} \in \bbbr^2$ is a corresponding
    2-D relative motion vector for the observer with index $k$ in
    relation to an observation from a reference view with index $\tilde{k}$,
  \end{itemize}
  see Figure~\ref{fig-pairgeom} for an illustration.
\end{itemize}
Under these types of geometric image transformations, define
the purely smoothed components of the receptive field responses
$L(x, t;\; s, \Sigma, \tau, v)$ and $L'(x', t';\; s', \Sigma', \tau', v')$
by convolving the respective image streams $f(x, t)$ and $f'(x', t')$
with velocity-adapted affine Gaussian kernels.
Then, these smoothed representations can be matched in such a way that
\begin{equation}
  \label{eq-joint-cov-prop-result-of-proof}
  L'(x', t';\; s', \Sigma', \tau', v') = L(x, t;\; s, \Sigma, \tau, v),
\end{equation}
provided that the receptive field parameters $(s, \Sigma, v, \tau)$ and
$(s', \Sigma', v', \tau')$ over the two domains are related in the following way
(Lindeberg \citeyear{Lin25-JMIV} Equations~(252)--(255))
\begin{align}
  \begin{split}
    \label{eq-s-transf-result}
    s' & = S_x^2 \, s,
  \end{split}\\
  \begin{split}
    \label{eq-Sigma-transf-result}
    \Sigma' & = A \, \Sigma \, A^{T},
  \end{split}\\
  \begin{split}
    \label{eq-tau-transf-result}
    \tau' & = S_t^2 \, \tau,
  \end{split}\\
  \begin{split}
    \label{eq-v-transf-result}    
    v' & = \frac{S_x}{S_t} (A \, v + u).
  \end{split}
\end{align}
Here, $s$ and $\tau$ denote the variance-based spatial and
temporal  parameters, corresponding to the squares of the
previous standard-deviation-based scale parameters.
Corresponding relations for the binocular/multi-view projection
model (\ref{eq-sc-aff-vel-transf-alt-obs-model}) are obtained
by replacing the matrix $A$ by $\tilde{B}^{(k)}$ and the
vector $u$ by $\tilde{u}^{(k)}$.
See Lindeberg (\citeyear{Lin25-JMIV,Lin26-JCompNeuroSci})
for a complementary theory for how to in addition match the
derivative-based components in the spatial or spatio-temporal
receptive field responses.

\section{Implications regarding orientation, direction and speed selectivity properties}

A direct consequence of the prediction about variabilities in the
shapes of spatial and spatio-temporal receptive fields according to
the generalized Gaussian derivative theory for visual receptive fields
is that these variabilities ought to also give rise to corresponding
variabilities in the shapes of orientation, direction and speed
selectivity curves.

In Lindeberg (\citeyear{Lin25-JCompNeurSci-spanelong}), we have shown
that such predictions regarding orientation selectivity properties are
consistent with the results of neurophysiological measurements by 
Nauhaus {\em et al.\/}\ (\citeyear{NauBenCarRin09-Neuron})
and Goris {\em et al.\/}\ (\citeyear{GorSimMov15-Neuron}).
  
Additionally, we have in Lindeberg (\citeyear{Lin25-arXiv-dirsel})
shown that such predictions regarding
(i)~speed selectivity properties are consistent with the results of
neurophysiological measurements by
Orban {\em et al.\/} (\citeyear{OrbKenNul86-JNeurPhys}), and
(ii)~direction selectivity properties are consistent with the results
of neurophysiological measurements by
Livingstone and Conway (\citeyear{LivCon03-JNeuroPhys}).

\section{Implementation in terms of neural computations}

Since the evolution property over the scale of the receptive field
responses are described by a diffusion equation of the form
(\ref{eq-vel-adapt-scsp-diff-eq}), it follows that this evolution process
can be discretized in terms of local computations, that can
be implemented in terms of neural-like computational units.
The smoothing with the time-causal limit kernel
(\ref{eq-time-caus-lim-kern}) can, in turn, be
implemented in terms of first-order integrators
coupled in cascade, according to the theory in
Lindeberg (\citeyear{Lin23-BICY}).

\section{Discussion}

We have described main components in the generalized Gaussian
derivative theory for visual receptive fields, concerning its
foundations, its use for formulating idealized models
for the receptive fields of simple cells, and its predictions about
variabilities in the shapes of the receptive fields.
These variabilities are intimately coupled to the influence of
geometric image transformations on the receptive field responses.

This theory has been formulated in an axiomatic way, building upon
assumptions that reflect symmetry properties in the image formation
process. It has been demonstrated to produce idealized receptive
field shapes that rather well resemble those recorded from
higher mammals by DeAngelis {\em et al.\/}\
(\citeyear{DeAngOhzFre95-TINS,deAngAnz04-VisNeuroSci}),
Conway and Livingstone (\citeyear{ConLiv06-JNeurSci}) and
Johnson {\em et al.\/}\ (\citeyear{JohHawSha08-JNeuroSci}).
This theory is also consistent with the variability
in the degree of elongation in receptive fields of mice
reported by Meshcherina {\em et al.\/} (\citeyear{MesAufLin20-bioRxiv}).

While Gabor filters have usually been more commonly used
for modelling visual receptive fields
(Marcelja \citeyear{Mar80-JOSA},
Jones and Palmer \citeyear{JonPal87a,JonPal87b},
Ringach \citeyear{Rin01-JNeuroPhys,Rin04-JPhys},
Citti and Sarti {\em et al.\/} \citeyear{CitSar06-JMIV},
Serre {\em et al.\/} \citeyear{SerWolBilRiePog07-PAMI},
Sarti {\em et al.\/} \citeyear{SarCitPet08-BICY},
Cocci {\em et al.\/} \citeyear{CocBarSar11-JOSA},
Barbieri {\em et al.\/} \citeyear{BarCitCocSar14-JMIV},
Baspinar {\em et al.\/} \citeyear{BasCitSar18-JMIV,BasSarCit20-MathNeuroSci}),
than the Gaussian derivative model
(Koenderink and van Doorn \citeyear{Koe84,KoeDoo87-BC,KoeDoo92-PAMI},
Young \citeyear{You87-SV},
Young {\em et al.\/}\ \citeyear{YouLesMey01-SV,YouLes01-SV},
Lindeberg \citeyear{Lin10-JMIV,Lin13-BICY,Lin21-Heliyon}),
conceptual advantages of the generalized Gaussian derivative theory
for visual receptive fields are that:
(i)~it is theoretically well-founded in terms of axiomatic
derivations, and
(ii)~has an associated theory of transformation properties under
geometric transformations, that is both more well-developed and 
mathematically more explicit compared to earlier such attempts
based on Gabor models.

Topics for future work include further investigating the validity of
using this theory for modelling biological vision, and of extending the
theory with more fine-grained modelling regarding the influence due
to illumination variations. The latter would then go beyond the
general invariance properties of receptive field responses over a
logarithmic brightness scale under local multiplicative illumination
and exposure variations, as described in Lindeberg
(\citeyear{Lin13-BICY,Lin21-Heliyon}).

\bibliographystyle{abbrvnat}

{\footnotesize
\bibliography{bib/defs,bib/tlmac}
}

\end{document}